\documentclass[aps,prb,twocolumn,superscriptaddress,showpacs,reprint]{revtex4-1}

\usepackage{graphicx, latexsym, verbatim}
\usepackage{amssymb, amsmath}
\usepackage[ansinew]{inputenc}
\usepackage{color}
\usepackage[english]{babel}

\begin{document}

\author{Troels Markussen}
\email{troels.markussen@gmail.com}
\affiliation{Center for Atomic-scale Materials Design (CAMD), Department of Physics,
Technical University of Denmark, DK-2800 Kgs. Lyngby, Denmark}

\title{Phonon Interference Effects in Molecular Junctions}


\date{\today}


\begin{abstract}
We study coherent phonon transport through organic, $\pi$-conjugated molecules. Using first principles calculations and Green's function methods, we find that the phonon transmission function in cross-conjugated molecules, like meta-connected benzene, exhibits destructive quantum interference features very analogous to those observed theoretically and experimentally for electron transport in similar molecules. The destructive interference features observed in four different cross-conjugated molecules significantly reduce the thermal conductance with respect to linear conjugated analogues. Such control of the thermal conductance by chemical modifications could be important for thermoelectric applications of molecular junctions.
\end{abstract}


\maketitle

\section{Introduction}
Recent experiments\cite{Losego2012,Ong2013,OBrien2013} have demonstrated that the thermal conductance of heterointerfaces between solids and monolayers of organic molecules, can be tuned by chemical modifications of the molecules. Progress in experimental scanning probe techniques has further enabled measurements of heat dissipation in single-molecule junctions\cite{Lee2013}. Such measurements and control of thermal transport at the nanoscale is important for several technological applications\cite{Cahill2003} and subject of intense research.  

Good thermal conductors and low interfacial thermal resistances are needed for e.g. computer processors in order to get the heat away. On the other hand, efficient thermoelectric materials require a low thermal conductance. Significant progress has been obtained to decrease the thermal conductance in nano-scale materials in order to increase the thermoelectric figure of merit $ZT=GS^2T/\kappa$ characterizing the efficiency of a thermoelectric material.\cite{DresselhausAdvMat2007}  Here $G$ is the electronic conductance, $S$ is the Seebeck coefficient, $T$ is temperature, and $\kappa$ is the thermal conductance with contributions from both electrons and phonons.

Molecular junctions in which a single molecule or a whole self-assembled monolayer (SAM) is sandwiched between solid state electrodes has been proposed as very promising candidates for thermoelectric applications\cite{Murphy2008,Bergfield2010,Nozaki2010,Dubi2011,Stadler2011} since the power factor $GS^2$ entering $ZT$ can be optimized through chemical engineering. Electronic conductance measurements of molecular junctions have been carried out in the last 15 years\cite{Reed1997}, and more recently it has become possible to measure the thermopower of single molecules\cite{ReddyScience2007}.
Furthermore, due to large difference in vibrational (phonon) frequencies in an organic molecule and in a typical solid, the thermal conductance of a molecular junction is believed to be low. This has indeed been experimentally demonstrated for both planar surfaces\cite{Wang2006,Wang2007} and very recently in nanoparticle networks\cite{Ong2013}. An other recent study demonstrated tuneability of the thermal conductance through molecular junctions by varying the anchoring groups connecting the molecular back-bone with gold electrodes\cite{Losego2012}. 

Several theoretical studies have considered phonon transport in molecular junctions using empirical/model potentials and Langevin methods\cite{Segal2003,ZurcherPRA1990,Hu2010} and atomistic Green's functions including anharmonic effects~\cite{Mingo2006}. Density functional theory (DFT) based methods within the harmonic approximation have been applied to study one-dimensional junctions with carbon nanotube\cite{JingTaoPRB2008}- and silicon nanowire-leads\cite{Nozaki2010} and for atomic Au chains\cite{EngelundPRB2009}. 

A number of recent measurements of \textit{electronic} transport have demonstrated large differences in conductance between linearly- and cross-conjugated aromatic molecules\cite{Fracasso2011,Hong2011,Guedon2012,Aradhya2012,Arroyo2013}. A molecule is linearly conjugated if it is possible to draw a path connecting the two ends which strictly alternate between single and double/triple bonds. A pathway in a $\pi$-conjugated molecule is cross-conjugated if it contains two subsequent single bonds and the (sp2 hybridized) carbon atom linking these single bonds is double-bonded to any group or atom in a third direction\cite{cross-conjugation}.  A molecule is called cross-conjugated, if all the pathways are cross-conjugated. Due to destructive quantum interference (QI) effects occurring in cross-conjugated molecules, these have several orders of magnitude lower electronic conductance as compared to linear-conjugated analogues showing no QI features in the relevant energy range around the Fermi energy. These experiments confirm a large number of theoretical studies predicting QI effects to be present in cross-conjugated molecules but not in linearly conjugated ones.\cite{Solomon2010,Solomon2008d,MarkussenNanoLett2010,Markussen2011}. 

In this paper we show that similar destructive QI effects have a significant impact on the phonon thermal conductance through molecular junctions. The calculated phonon transmission functions through several cross-conjugated aromatic molecules (e.g. meta-connected benzene) all display clear QI effects. This leads to differences in the room temperature thermal conductance of factors 2-5 when comparing to linear conjugated junctions (e.g. para-connected benzene). Although these ratios depend on the specific details of the electrode material and the molecule-electrode coupling, our findings are robust, predicting significant effects of conjugation pattern on the phonon thermal conductance. Phonon interference effects have recently been studied theoretically for alkane SAM interfaces as function of SAM thickness\cite{Hu2010,Zhang2011}. However, the observed Fabry-Perot like interference effects are constructive and the linear alkanes do not lead to complete destructive QI as is the case for the cross-conjugate aromatic molecules studied here.

In the remaining parts of the paper we first describe our computational methods. This is followed by a results section where we first consider a prototypical model in the case of benzene described with a simple model. We next consider the more realistic OPE3 molecule as well as other molecules using DFT and empirical potential methods. We end up with a discussion and conclusion.

\section{Methods} \label{method}
In this work we limit ourself to the harmonic approximation, thus neglecting anharmonic phonon-phonon scattering. This is a reasonable approximation since anharmonic scattering is of limited importance due to the shortness of the molecular junction\cite{Segal2003}.
In the harmonic approximation, the phononic system is fully determined by the force constant matrix, $\mathbf{K}$, which contains the spring constants between atoms $I$ and $J$ in directions $\mu$ and $\nu$. We calculate $\mathbf{K}$ using either density functional theory (DFT)  or the semi-empirical Brenner force field\cite{Brenner1990} as implemented in the 'general utility lattice program' (GULP)\cite{gulp}. For the DFT calculations we use GPAW\cite{gpaw-review}, which is an electronic structure code based on the projector-augmented wave method. The DFT calculations are performed with a double zeta polarized atomic orbital basis set and the exchange correlation potential described by the Perdew-Burke-Ernzerhof (PBE) functional\cite{PBE}. 

After an initial relaxation, each atom, $I$, is displaced by $Q_{I \mu}$ in direction $\mu=\{x,y,z\}$ to obtain the forces, $F_{J\nu}(Q_{I\mu})$, on atom $J\neq I$ in direction $\nu$. The structure is initially relaxed with a maximum residual force of 0.01 eV/\AA. The ions are displaces by $Q_{I \mu}=\pm 0.05\,$\AA. The force constant
matrix, $\mathbf{K}$, is then found by finite differences
\begin{equation}
K_{I\mu,J\nu}=\frac{\partial^2E}{\partial R_{I\mu}\partial R_{J\nu}}=\frac{F_{J\nu}(Q_{I\mu})-F_{J\nu}(-Q_{I\mu})}{2Q_{I\mu}},
\end{equation}
with $E$ being the total energy. The intra-atomic elements are calculated by imposing momentum conservation, such that $K_{I\mu,I\nu}=-\sum_{K\neq I}K_{I\mu,K\nu}$. From the force constant matrix we obtain the dynamical matrix, $\mathbf{D}$, with $D_{I\mu,J\nu}=K_{I\mu,J\nu}/\sqrt{M_IM_J}$, where $M_{I}$ is the mass of atom $I$. The phonon eigen-frequencies, $\omega_i$ and eigenmodes, $u_i$ are obtained from the equation of motion 
\begin{eqnarray}
\mathbf{D}u_i=\omega_i^2u_i.
\end{eqnarray}

The phonon transmission function is calculated as\cite{Mingo2006}
\begin{eqnarray}
\mathcal{T}(\omega) = {\rm Tr}\left[G^r(\omega)\Gamma_L(\omega)G^a(\omega)\Gamma_R(\omega)\right],
\end{eqnarray}
where $G^{r}(\omega)=(\omega^2-\mathbf{D}-\Sigma_L-\Sigma_R)^{-1}$ is the retarded (advanced) Green's function, and $\Gamma_{L,R}(\omega)=i(\Sigma^r_{L,R}(\omega)-\Sigma_{L,R}^a(\omega))$ describes the coupling to the left and right leads expressed in terms of the lead self-energies $\Sigma_{L,R}(\omega)$. We have used the Brenner potential\cite{Brenner1990} to simulate silicon nanowire (SiNW) and graphene nanoribbon (GNR) leads. In this case we get an atomistic description of the leads and the self-energies. In addition to this, we also take a simpler approach, where we compute the dynamical matrix for a \textit{free} molecule with DFT, and following Mingo\cite{Mingo2006} we model the lead surface density of states using an analytical form $\rho(\omega)=-\frac{1}{\pi}{\rm Im}g^0(\omega)=\frac{3\omega}{2\omega_D^3}\Theta(\omega_D-\omega)$ for the imaginary part of the surface Green's function $g^0(\omega)$. The corresponding real part is obtained from a Hilbert transformation. We thus characterize the bare leads with a single parameter, $\omega_D$. Unless otherwise noticed, we use $\hbar\omega_D=70meV$, thus resembling a phonon density of states similar to that of e.g. silicon. We also use the atomic mass of silicon for the lead mass, $M_L$. A plot of the real- and imaginary part of $g^0(\omega)$ is provided in Appendix \ref{Gsurf}. The surface Green's functions $g^0(\omega)$ do not include the coupling to the molecule. To include this, we use the Dyson equation\cite{Mingo2006} to get 
\begin{equation}
g(\omega)=g^0(\omega)[1+\tilde{\gamma} g^0(\omega)]^{-1},
\end{equation}
where $\tilde{\gamma}=\gamma/\sqrt{M_cM_L}$ is the mass-scaled coupling force constant between the leads and the molecule, and $M_c$ and $M_L$ are the carbon- and lead atomic masses. We characterize the molecule-lead coupling with two adjustable parameters describing the out-of-plane motion ($\gamma_z$) and the in-plane motion ($\gamma_{xy}$). In all calculations shown below we use $\gamma_z=-4.0\,$eV/\AA$^2$. For some calculations we only consider out-of-plane ($z$-direction) motion, in which case we set $\gamma_{xy}=0$. Otherwise we use $\gamma_{xy}=\gamma_z$.
The lead self-energy on the molecule is finally 
\begin{equation}
\Sigma_\nu(\omega) = \gamma_\nu^2g(\omega),
\end{equation}
for each degree of freedom $\nu=\{x,y,z\}$. Assuming that site 1 and N on the molecule are connected to the left and right leads, we have for the self-energy matrices $[\Sigma_L(\omega)]_{1\nu,1\nu}=\Sigma_\nu(\omega)$ and $[\Sigma_R(\omega)]_{N\nu,N\nu}=\Sigma_\nu(\omega)$ with zeros elsewhere. After coupling the molecule and the leads, the force constant matrix from the free molecule is corrected in order to fulfill momentum conservation. Assuming again that site 1 ($N$) are connected to the left (right) leads  we change on-site elements $K_{1(N)\nu,1(N)\nu}\rightarrow K_{1(N)\nu,1(N)\nu}-\gamma_\nu$ with $\nu=\{x,y,z\}$.
While the simplified description of the leads does not capture all details of a real surface, the focus in this paper is on the intrinsic phonon transport properties of the molecule, which are robust against the specific coupling to the leads, as shown below.

From the phonon transmission function $\mathcal{T}(\omega)$, the phonon thermal conductance is calculated as 
\begin{equation}
\kappa_{ph}(T) = \frac{\hbar^2}{2\pi k_B T^2}\int_{0}^\infty\mathrm{d}\omega\,\omega^2\,\mathcal{T}(\omega)\,\frac{e^{\hbar\omega/k_BT}}{(e^{\hbar\omega/k_BT}-1)^2}, \label{ThermalConductance}
\end{equation}
where $T$ is the average temperature of the left- and right leads.

\section{Results}
\subsection{Benzene - simple model}
\begin{figure}[htb!]
\includegraphics[width=\columnwidth]{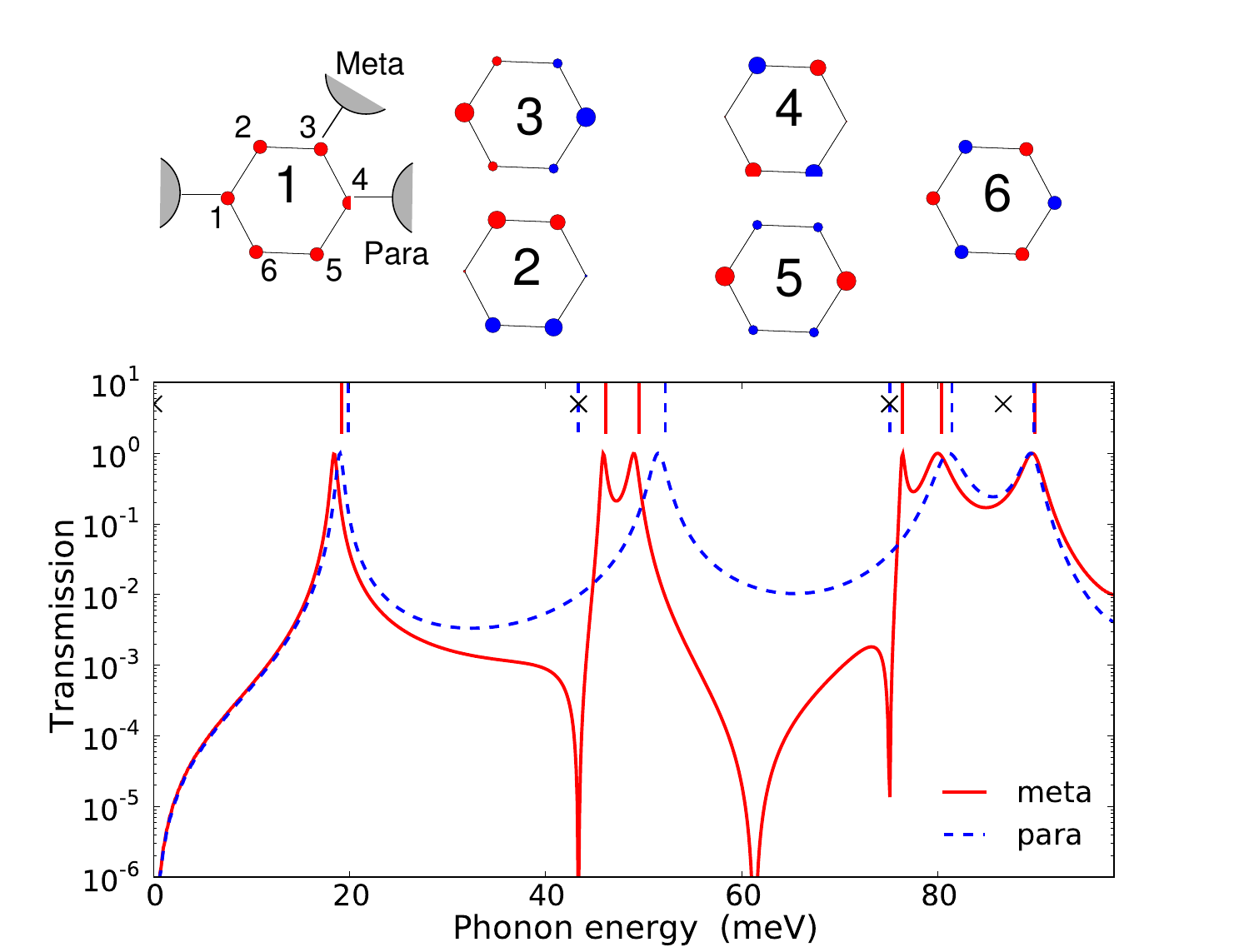}
\caption{Phonon transmission function through meta- (solid red) and para-connected (dashed blue) benzene calculated with a simplified nearest-neighbour model considering only out-of-plane carbon motions, and using a lead cut-off frequency $\hbar\omega_D=100\,$meV. The top part shows the symmetries of the phonon modes for the free molecule and illustrates the meta- and para connections. The energies of the free-molecule modes are indicated with crosses in the lower panel. The energies of the modes of the molecules coupled to leads are indicated with vertical lines.}
\label{benzene:fig}
\end{figure}

The earliest experimental evidences for QI in electron transport were concerned with comparing para- and meta-connected benzene molecules\cite{Patoux1997,Mayor2003b} (see Fig. \ref{benzene:fig}), which has also been analyzed theoretically( e.g. Refs. \cite{Sautet1988,Hansen2009}). We begin our analysis with also considering benzene. Initially we only consider out-of-plane vibrations and we assume only nearest neighbor interactions with force constants $K_{Iz,Jz}=k=-5.4\,$eV/\AA$^2$. Also, for simplicity we initially neglect the hydrogen atoms. Although these simplifications greatly reduce the number of vibrational modes, we note that inclusion of hydrogen mainly leads to a minor down-shift of the out-of-plane modes together with introduction of very high-frequency modes, which are far above the lead vibrational spectrum and hence do not contribute to the transport. Moreover, the out-of-plane vibrations in aromatic molecules are much softer than the in-plane vibrations. This means that the out-of-plane modes typically have energies below 100 meV, while the in-plane spectrum is mainly above 100 meV.

In the simplified case of only out-of-plane carbon motions, the dynamical matrix has a form very similar to the electronic Hamiltonian within a $\pi$-orbital model. The electronic conductance is determined by the transmission function at the Fermi level, which will usually be close to the $\pi$ orbital on-site energies, $\varepsilon_\pi$. It can be shown\cite{MarkussenNanoLett2010} that the electronic transmission function of meta-connected benzene has a node at energy $E=\varepsilon_\pi$, while para-benzene does not. This is the reason for the experimentally observed lower electronic conductance for meta-benzene\cite{Mayor2003b,Arroyo2013}.

In analogy with the electronic transmission node at $E=\varepsilon_\pi$ we expect a phonon transmission node, when $\omega^2=D_{II}$, i.e. when the squared frequency equals the diagonal elements of the dynamical matrix. In Appendix \ref{graphical} we give a proof if this result. This leads to an expected phonon transmission node at phonon energy $E_{ph}=\hbar\sqrt{-2k/M}=61.3\,$meV, where $M$ is the carbon atomic mass and where we use the sum rule $K_{II}=-\sum_{J\neq I}K_{IJ}$. Figure \ref{benzene:fig} (bottom) shows the phonon transmission functions of meta- and para-connected benzene calculated for the simple model with analytical self-energies.	In agreement with the expectation we observe a clear transmission dip around $E_{ph}=61\,$meV due to destructive phonon interference, whereas no such interference features are observed for the para-connection. In addition to the main QI transmission node for meta-benzene, two additional, more narrow transmission nodes are observed. By applying a graphical scheme for analyzing transmission nodes\cite{Markussen2011} (see Appendix \ref{graphical}), one can readily derive that the additional transmission nodes occur at $E_{ph}=\hbar\sqrt{-2k/M \pm|k|/M}$ giving 43.4 eV and 75.1 eV in agreement with the numerical result. As shown in Appendix \ref{graphical}, transmission node energies are independent on the strength of the molecule-lead coupling.


In the top row of Fig. \ref{benzene:fig} we illustrate the six out-of-plane phonon modes. These modes have been calculated for the free molecules, i.e. without the coupling to the electrodes. The modes have energies $E_1=0.0\,$eV, $E_{2,3}=43.4\,$eV, $E_{4,5}=75.1\,$eV, and $E_{6}=86.7\,$eV are indicated with crosses in lower part of Fig. \ref{benzene:fig}.  The size of the circles indicate the amplitude of the displacement and the color indicate the phase. The lowest energy mode is just a rigid displacement of all atoms and has zero energy for the free molecule.  When coupling to the leads is taken into account, the modes are shifted upward in energy, but essentially preserve their shapes (see also Appendix \ref{graphical}). The coupled-mode energies are indicated with vertical lines in the transmission plot in Fig. \ref{benzene:fig}. The most pronounced energy shift is seen for Mode 1, which is shifted from zero energy to $E_1=19\,$eV. Modes 2 and 3 are degenerate for the free molecule as are modes 4 and 5. However, these degeneracies are lifted when the lead-coupling is taken into account. Notice that modes 2 and 5 have zero weight on the atoms connecting to the leads in the para configuration. This means that are not energy shifted when coupled to leads. Further, they do not contribute to the transport at all, and in the para-connection we thus only see transport through modes 1,3,4, and 6 each giving a transmission peak with a maximum value of 1. In the meta-configuration all modes are coupled to the electrodes and we observe six distinct peaks in the transmission  function.

\begin{figure}[htb!]
\includegraphics[width=0.9\columnwidth]{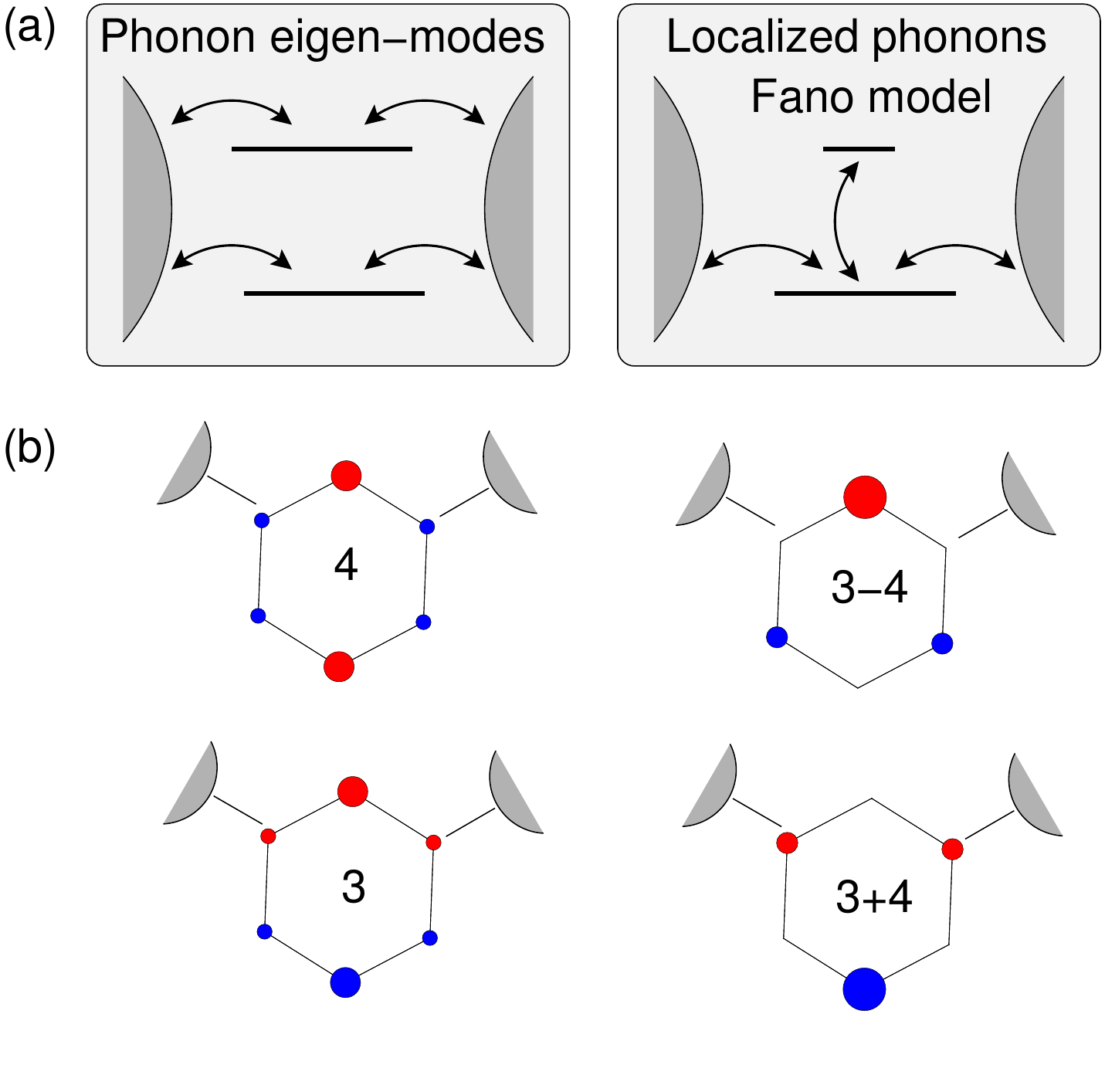}
\caption{Pairs of phonon eigen-modes (shown in left column) can be superposed to form localized phonon modes (right column). The topology of the localized phonon model, which is schematically shown in panel (a), right, closely resembles a Fano resonance with a localized mode coupled to a continuum.  Panel (b) shows eigen-modes 3 and 4 (left part) of benzene. In the right part the two modes are either added (3+4) or subtracted (3-4).}
\label{mode-fig}
\end{figure}

Additional insight into the central QI transmission node at $E_{ph}=61.3\,$meV can be obtained from an analysis of the eigen-modes of the free molecule. Figure \ref{mode-fig} (b) show to the left the eigen-modes 3 and 4 of benzene (within the simplified nearest-neighbor model). Both modes are connected to leads in a similar manner (both with equal phases to the left and right contact) and it may not at first be obvious that this leads to QI induced transmission nodes. This situation is schematically shown in panel (a) left. If we form linear combinations of the two modes we may obtain localized phonon modes (LPM) as shown in the right part of panel (b). This representation is just as general as the eigen-mode basis and may be easier for interpretation of the QI transmission nodes. Now we have a situation, where one of the LMPs (3+4) is delocalized and connected to both leads, whereas the other is localized and decoupled from the leads, but coupled to the delocalized mode, as shown schematically in panel (a) right. This LPM topology closely resembles the situation of a Fano resonance\cite{Miroshnichenko2010} where a localized state couples to a continuum. Similar to electronic transport it will always lead to a transmission node at the vibrational frequency of the localized phonon mode\cite{Stadler2011}, $\hbar\omega_{3-4}=\hbar\sqrt{(\omega_{3}^2+\omega_{4}^2)/2}=61.3\,$meV. Since the localized mode does not couple to the leads, the energy is unaffected by the lead coupling and the transmission node energy is independent of the molecule-lead coupling, as discussed above.

We notice that when eigen-modes 2+5 and 1+6 are combined to form LPMs we also obtain the Fano-like model with a transmission zero at the same energy as for mode 3 and 4.


\subsection{OPE3 - DFT based calculations}

\begin{figure}[htb!]
\includegraphics[width=\columnwidth]{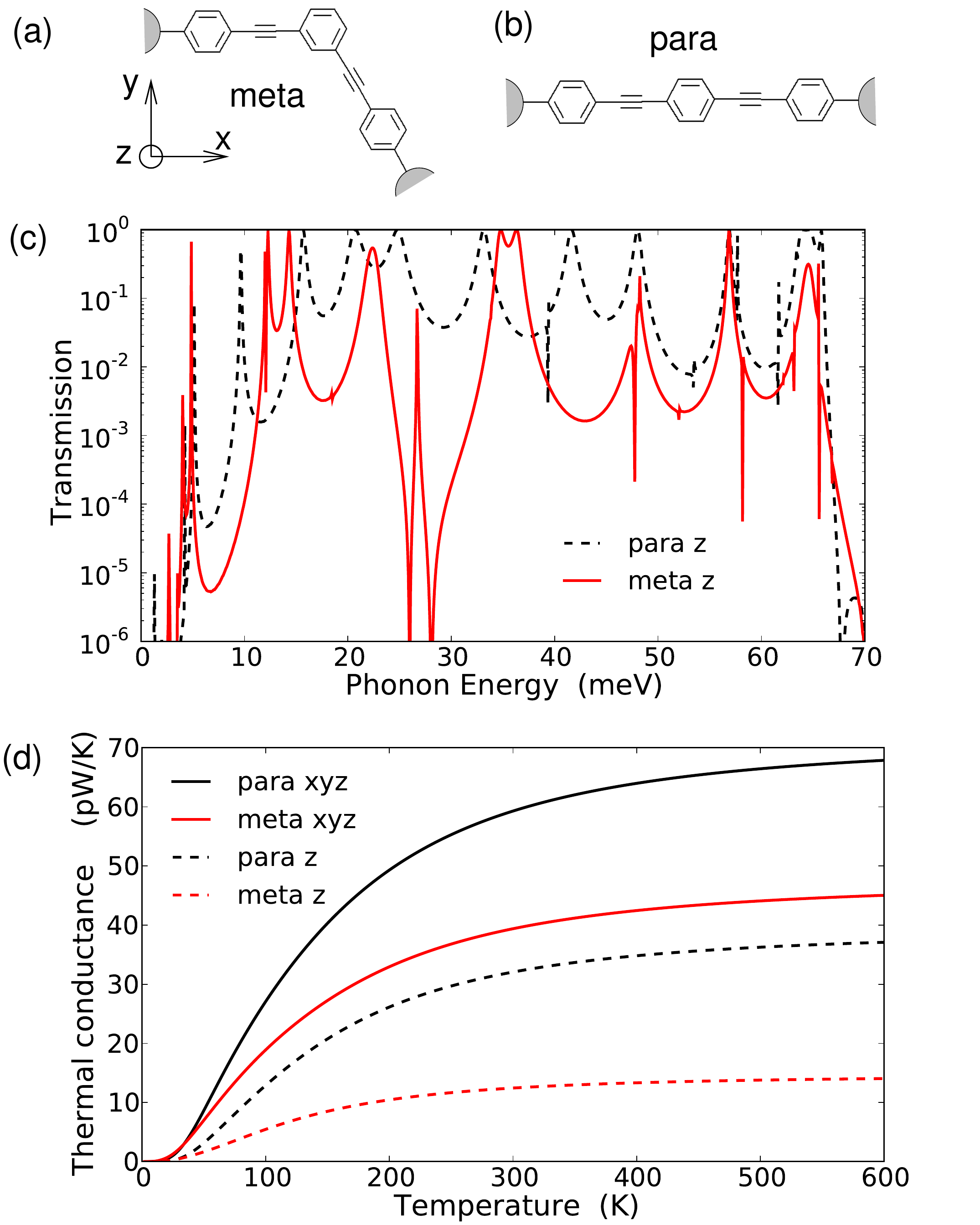}
\caption{Structure of meta- (a) and para-OPE3 (b) and phonon transmission functions (c) for meta- (solid red) and para-OPE3 (dashed black) calculated from only out-of-plane ($z-$direction) motion. Panel (d) show the thermal conductance vs. temperature in the two cases of only out-of-plane motion (dashed lines) and inclusion of all modes, i.e. including motion in all  $(x,\,y,\,z)$-directions (solid lines).}
\label{thio-benzene:fig}
\end{figure}

In order to address the generality of the results obtained from the simple model calculations we consider in Fig. \ref{thio-benzene:fig}  two oligo(phenylene–ethynylene) (OPE3) molecules; one with the central benzene in a meta-connection (a) and the other in a para-configuration (b). The dynamical matrix of the free molecules are now calculated with DFT as described above. Panel (c) shows the transmission functions for the out-of-plane motion. In qualitative agreement with the simple model calculations in Fig. \ref{benzene:fig} we again observe several QI induced transmission nodes for the meta- (solid red) but not for the para-configuration (dashed black). 

\begin{figure}[htb!]
\includegraphics[width=\columnwidth]{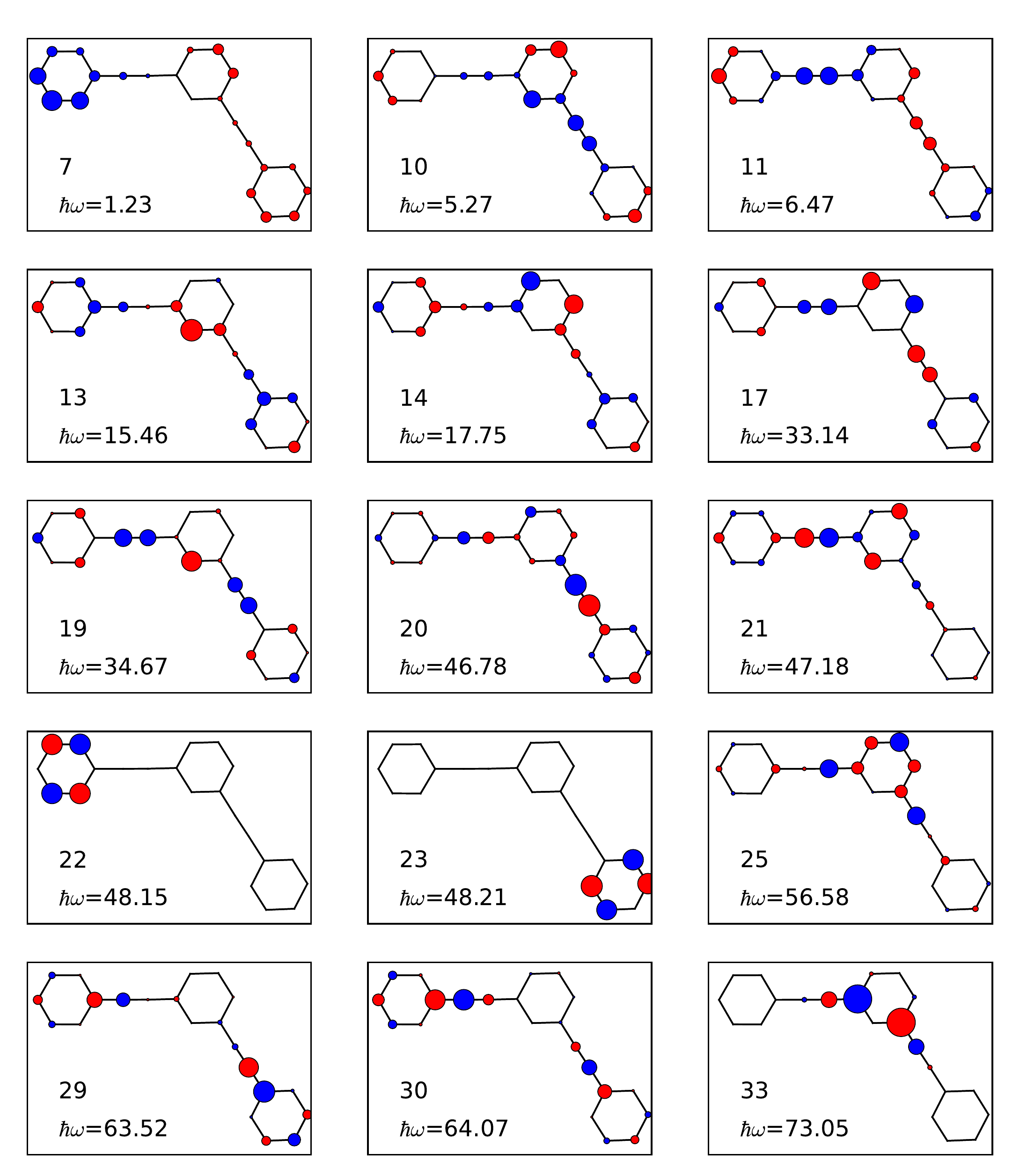}
\caption{Out-of-plane modes for meta-OPE3 within the transport energy window. For each mode we indicate the mode index and the eigen-energy (in units of meV).}
\label{ope3-z-modes}
\end{figure}

\begin{figure}[htb!]
\includegraphics[width=0.9\columnwidth]{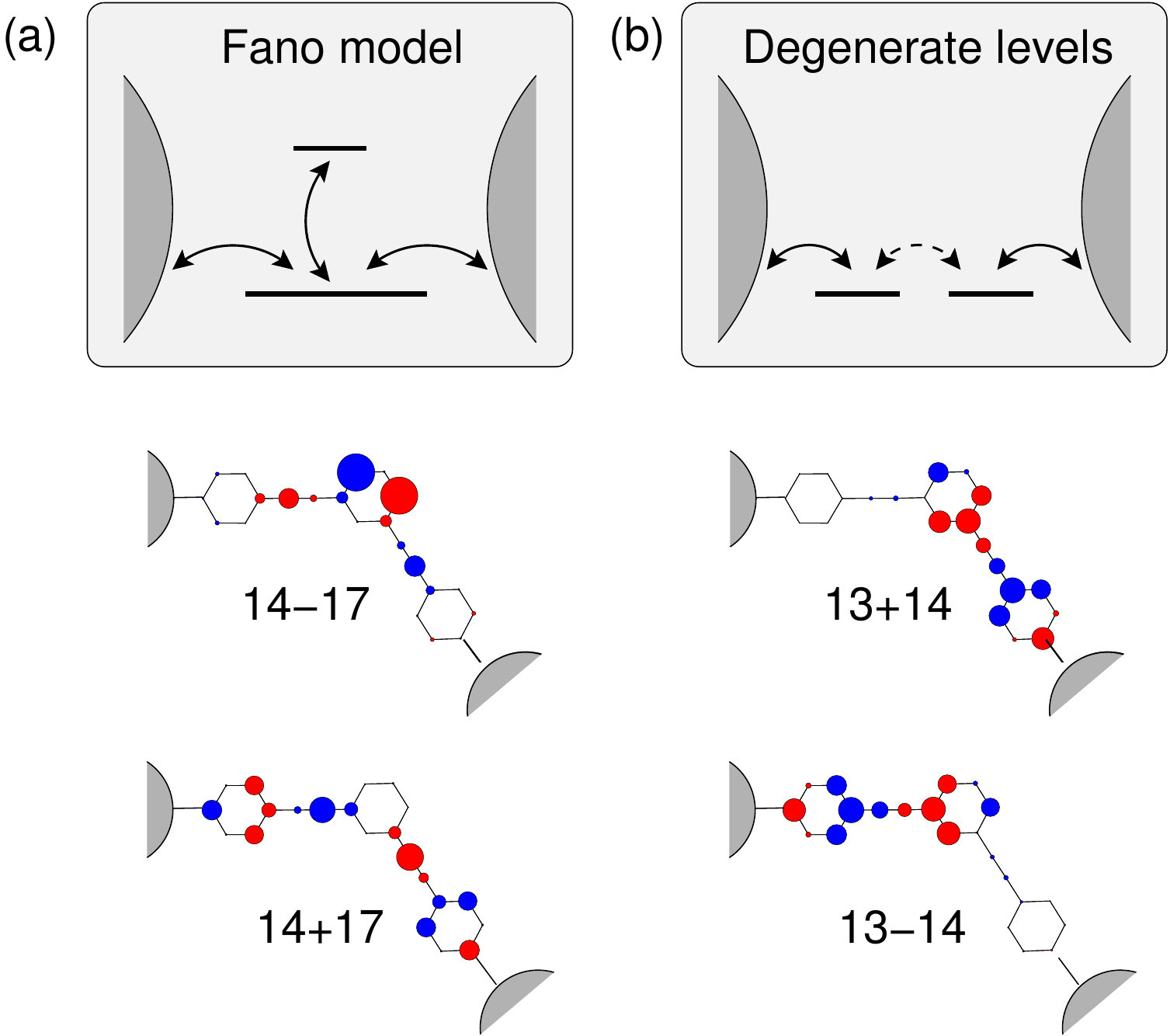}
\caption{Localized phonon modes formed by combining eigen-modes. The Fano-model (a) leads to a transmission zero at the energy of the localized, decoupled, mode ($E_{ph}=26.6\,$meV). The degenerate levels model leads to a reduced transmission peak at $E_{ph}=22\,$meV due to a weak coupling between the left- and right localized states.}
\label{ope3-two-levels}
\end{figure}

By inspecting the out-of-plane modes of meta-OPE3 shown in Fig. \ref{ope3-z-modes} we find that the combination of eigen modes leading to a Fano-model, discussed above for benzene, can also be applied to the meta-connected OPE3 (the out-of-plane modes for para-OPE3 are shown in Appendix \ref{para-modes}). Here we also find pairs of eigen-modes corresponding to energies on either side of a transmission node, which form a delocalized LPM connected to both leads and a localized LPM being dis-connected from the leads. An example is shown in Fig. \ref{ope3-two-levels} (a) where modes 14 and 17 are responsible for the transmission node around $E=25-27\,$meV. The eigen-modes 14 and 17 of meta-OPE3 have energies $\hbar\omega_{14}=17.8\,$meV and $\hbar\omega_{17}=33.1\,$meV. These two modes are successive when considering modes containing only out-of plane motion and are the two modes closest in energy below/above the QI transmission node -- see also Fig. \ref{ope3-z-modes}. When forming linear combinations of the two modes we obtain localized phonon modes (LPM) as shown in Fig. \ref{ope3-two-levels} (a). Both LPMs have a local (on-site) frequency of $\hbar\omega_{14\pm17}=\hbar\sqrt{(\omega_{14}^2+\omega_{17}^2)/2}=26.6\,$meV. In similarity with the meta-benzene case, the LPM topology for meta-OPE3 resembles a Fano model with a transmission node at the energy of the localized mode. In this particular case, modes 14 and 17 are thus responsible for the transmission node around $E=25-27\,$meV. From Fig. \ref{ope3-z-modes} we observe that modes 13 and 19 have similar symmetries as modes 14 and 17. Performing the same analysis we find that modes 13 and 19 also lead to a Fano-model with expected transmission node at energy $E_{ph}=26.8\,$meV, i.e. very close to the result for modes 14 and 17. The narrow peak in the transmission spectrum at $E=26\,$meV is due a mode with primarily in-plane motion character, not shown in Fig. \ref{ope3-z-modes}. The transmission dip seen around $E_{ph}=47\,$meV results from an interplay between the four nearly degenerate modes 20--23.

In addition to the phonon interference effects leading to transmission nodes, another type of interference effect due to quasi degenerate modes also play a role\cite{Hartle2011} . To illustrate this, consider the quasi-degenerate modes 13 and 14 in Fig. \ref{ope3-z-modes}. The LPMs obtained from these modes are shown in Fig. \ref{ope3-two-levels} (b). The left- and right-localized LPMs are only weakly coupled with a coupling strength proportional to the energy separation between the eigen modes. This is the explanation of the reduced transmission peak around $E_{ph}=22\,$meV for meta-OPE3 contributing to the reduced thermal conductance of meta-OPE3. Note that due to the molecule-lead coupling, modes 13 and 14 are shifted from their free values of $15-17\,$meV to $\sim22\,$meV. Similar interference effects due to quasi-degenerate electronic states has recently been analyzed theoretically for electron transport\cite{Hartle2011}.

In Fig. \ref{thio-benzene:fig} panel (d) we show the thermal conductance vs. temperature for the two OPE3 molecules. The solid curves are calculated with molecule-electrode coupling for motions in all directions $(xyz)$ while the dashed curves are obtained when only the out-of-plane ($z-$ direction) motion is considered. In either case we observe a clear difference between para- and meta-connections, with the former having 2 times higher conductance. When only the out-of plane motion is considered the para to meta ratio is close to 3.

\subsection{OPE3 with SiNW- and GNR leads}

\begin{figure*}[htb!]
\includegraphics[width=\textwidth]{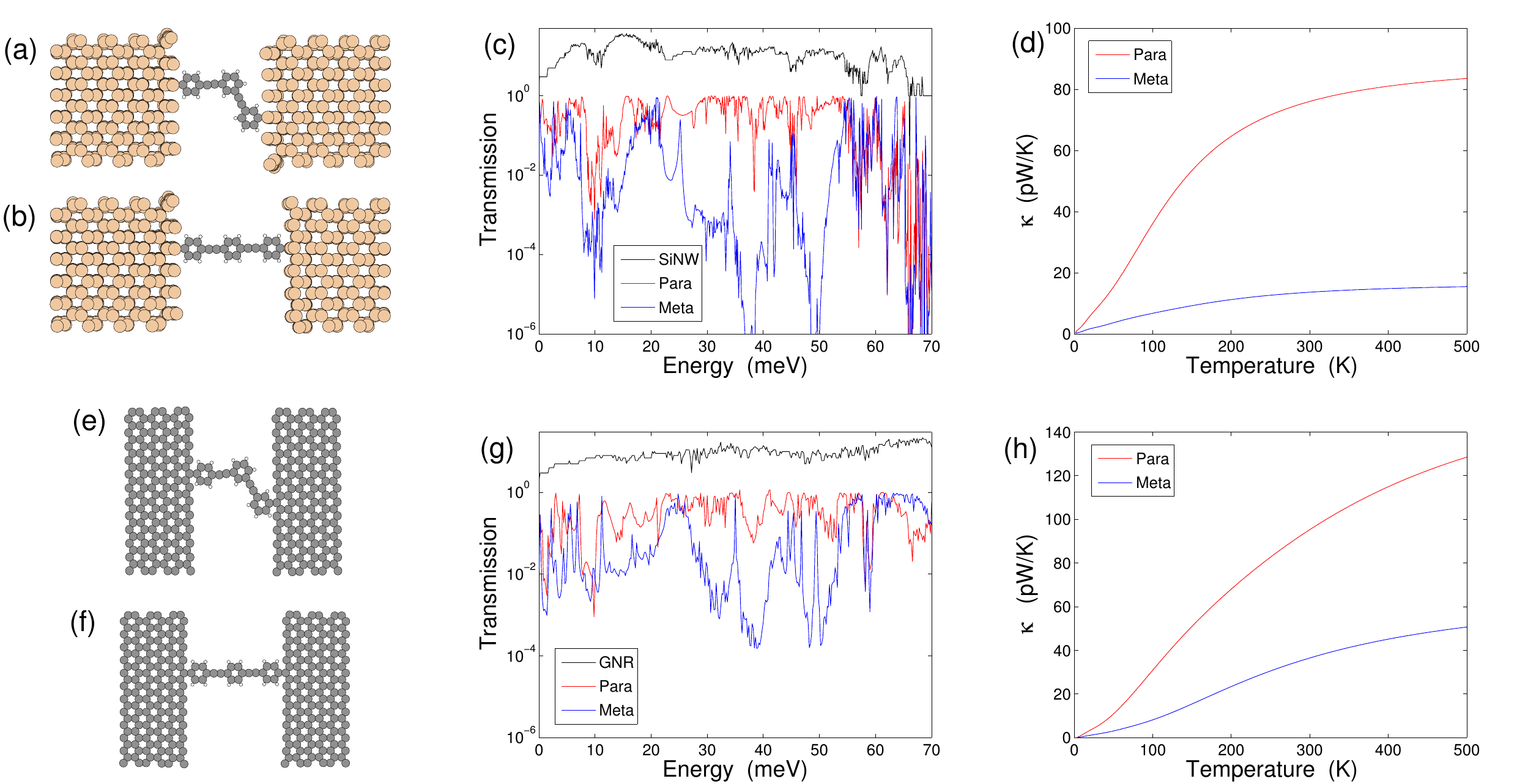}
\caption{Panels (a) and (b) show the atomic structures of meta- and para-OPE3 connected to SiNW leads, while panels (e) and (f) show the structures with GNR leads. Panels (c) and (g) show the phonon transmission functions of meta (blue), para (red), and pure SiNW/GNR (black). The corresponding thermal conductances are shown in panels (d) and (h).}
\label{si-gnr-elec-fig}
\end{figure*}

In order to test the robustness of the results obtained with the analytical self-energies we have performed calculations with more realistic leads. In Fig. \ref{si-gnr-elec-fig} we show the atomic structure of the two OPE3 molecules connected to silicon nanowire (SiNW) leads (a) and (b) and to graphene nano-ribbon (GNR) leads, panels (e) and (f). The dynamical matrices of both the leads and the molecules have been calculated with the empirical Brenner potential\cite{Brenner1990} after an initial relaxation as described above. Panel (c)/(g) show the phonon transmission through the pristine SiNW/GNR (black) and through the two OPE3 molecules,  including all the phonon modes. In clear accordance with the DFT based calculations using analytical self-energies (Fig.  \ref{thio-benzene:fig} (c)) we again observe clear transmission dips for meta-OPE3 but not for para-OPE3. The corresponding thermal conductances are shown in panels (d) and (h), again showing a very substantial difference between meta- and para configurations. While the results obtained with the SiNW and GNR leads show that the absolute values of the conductances depend on the details of the leads and on the molecule-lead coupling, it is also clear that the general trends obtained with the simple analytical lead self-energies are robust against such details. In particular we observe a significant difference in thermal conductance between para- and meta OPE3 in all cases due to destructive QI effects in meta-OPE3.

\subsection{More molecules}

\begin{figure}[htb!]
\includegraphics[width=\columnwidth]{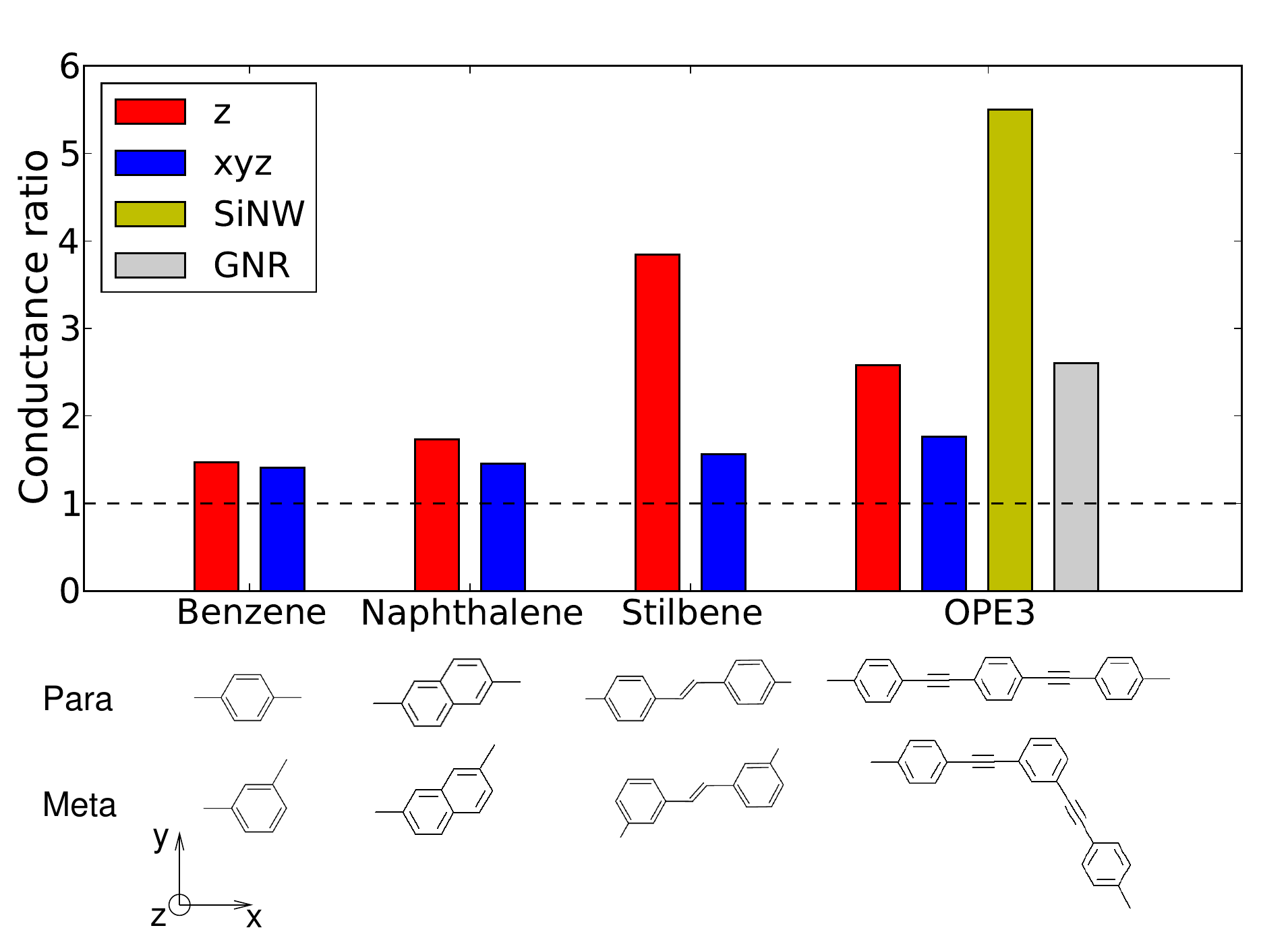}
\caption{Ratios between thermal conductances of the para- and meta-connected molecules shown in the bottom. Results are shown in the cases of only out-of-plane motion ($z$-direction, red bars) and inclusion of all modes ($xyz$-directions, blue bars). For OPE3 we also show the ratios obtained with SiNW (green bar) and GNR leads (gray bar) using the Brenner potential. The temperature is 300 K. }
\label{ratios-fig}
\end{figure}

To further address the general influence of conjugation pattern on the thermal conductance we have studied three additional pairs of molecular junctions. Figure \ref{ratios-fig} shows the thermal conductance ratios between para- and meta-configurations of the four pairs of molecules shown in the bottom rows. For all molecules we observe a pronounced effect of conjugation when only the out-of-plane ($z$-direction) motion is considered (red bars). When the in-plane ($x-$ and $y$-directions) motion is included (blue bars), the conductance ratios are decreased, but still show sizable effects of conjugation. For OPE3 we also show ratios obtained with SiNW (green bar) and GNR leads (gray bar) using the empirical Brenner potential. The fact that we also obtain high conductance rations between para- and meta-OPE3 using fully atomistic descriptions of the SiNW- and GNR-leads supports the use of the simple analytical lead self-energies. Furthermore, it clearly underlines that the phonon interference effect is an intrinsic property of meta-OPE3 and not due to a particular junction configuration.

\section{Discussion}
The results presented above show that the variations in thermal conductance with conjugation pattern follow the expected variations in the electronic conductance. For both electron and phonon transport the difference between meta- and para-configurations are caused by QI induced transmission nodes present in the former case but absent in the latter. While variations in the electronic conductance between linearly conjugated molecules (para) and cross-conjugated ones (meta) may be 1-3 orders of magnitude\cite{Fracasso2011,Hong2011,Guedon2012,Aradhya2012,Arroyo2013}, the corresponding differences for the phonon thermal conductance calculated here is rather a factor of $\sim 2-5$. The difference between electron- and phonon conductances is due to the differences in the Fermi- and Bose-Einstein distribution functions determining the respective occupations. Electron transport is determined by the transmission function in a narrow energy window around the Fermi level, $E_F$, and is thus very sensitive to potential QI effects close to $E_F$. On the contrary, the phonon thermal conductance depends on the transmission function in a large frequency range involving many phonon modes, where only some of them leads to QI effects.
We notice that recent experiments observed a factor of $\sim 2$ differences in thermal conductance between strongly (covalently) sulfur-gold bonded molecules and weak van-der Waals bonded CH$_3$-gold bonded contacts. However, changes in anchoring groups often leads to orders of magnitude changes in the electronic conductance\cite{Chen2006}. Although the thermal conductance variations due to variations in conjugation pattern are much smaller than the corresponding electronic conductance variations, the influence of conjugation seems to be at least as important as the influence of molecule-lead coupling, and should be well within an experimentally observable range. 

In a measurement of the thermal conductance there will be contributions from both phonons and electrons, while in this work we have only considered the phonon part. In order to make an accurate comparison of the phonon- and electron contributions one should perform calculations for both the phonon- and electron transmission functions using the same atomistic description. This is beyond the scope of this paper. However, an estimate of the electronic contribution to the thermal conductance can be made using the Wiedemann-Franz law $\kappa_{el}=LTG_{el}$, where $T$ is the temperature, $L=\pi^2k_B^2/(3e^2)$ is the Lorenz number, and $G_{el}$ is the electronic conductance. For molecular junctions typical values of $G_{el}$ are in the range $10^{-4}\,G_0$ to 0.1 $G_0$, where $G_0=2e^2/h$ is the electronic conductance quantum. Large conductance values are typically obtained for short, conjugated molecules. Cross-conjugated molecules have 10-1000 times lower conductance\cite{Fracasso2011,Hong2011,Guedon2012,Aradhya2012,Arroyo2013}.  We emphasize that theses values depend strongly on the specific molecule, the electrode material as well as the anchor group between molecule and electrode. The corresponding electronic thermal conductance values are in the range $\kappa_{el}=0.1-110$ pW/K. Comparing with the phonon thermal conductance values obtained above ($\sim 10-100\,$pW/K), we estimate that the phononic contribution to $\kappa$ is likely to be the dominant one for the cross-conjugated molecules (meta-connection), while the electronic and phononic contributions to $\kappa$ could be similar in magnitude in the linear conjugated molecules (para-connection). In all circumstances, inclusion of the electronic contributions to $\kappa$ will increase the ratio between linear- and conjugated molecules.

\section{Conclusion}
In conclusion, we have studied theoretically the influence of conjugation pattern on the phonon thermal conductance in molecular junctions. Similar to electronic transport we observe nodes in the transmission function due to destructive interference effects. The transmission nodes are observed for the cross-conjugated (meta-connection) molecular junctions  resulting in significantly lower thermal conductance when compared with the linear-conjugated  analogue junctions (para-connection). We generally observe a factor 2-5 reduction of the thermal conductance in the meta configurations. These findings might be important for thermal management at the nanoscale and in particular for thermoelectric applications where one seeks a low thermal conductance.

\begin{acknowledgements}
I greatly acknowledge financial support from the Danish Council for Independent Research, FTP Grants No. 11-104592 and No. 11-120938.
\end{acknowledgements}


\appendix
\section{Surface Green's function} \label{Gsurf}
\begin{figure}[htb!]
\includegraphics[width=0.8\columnwidth]{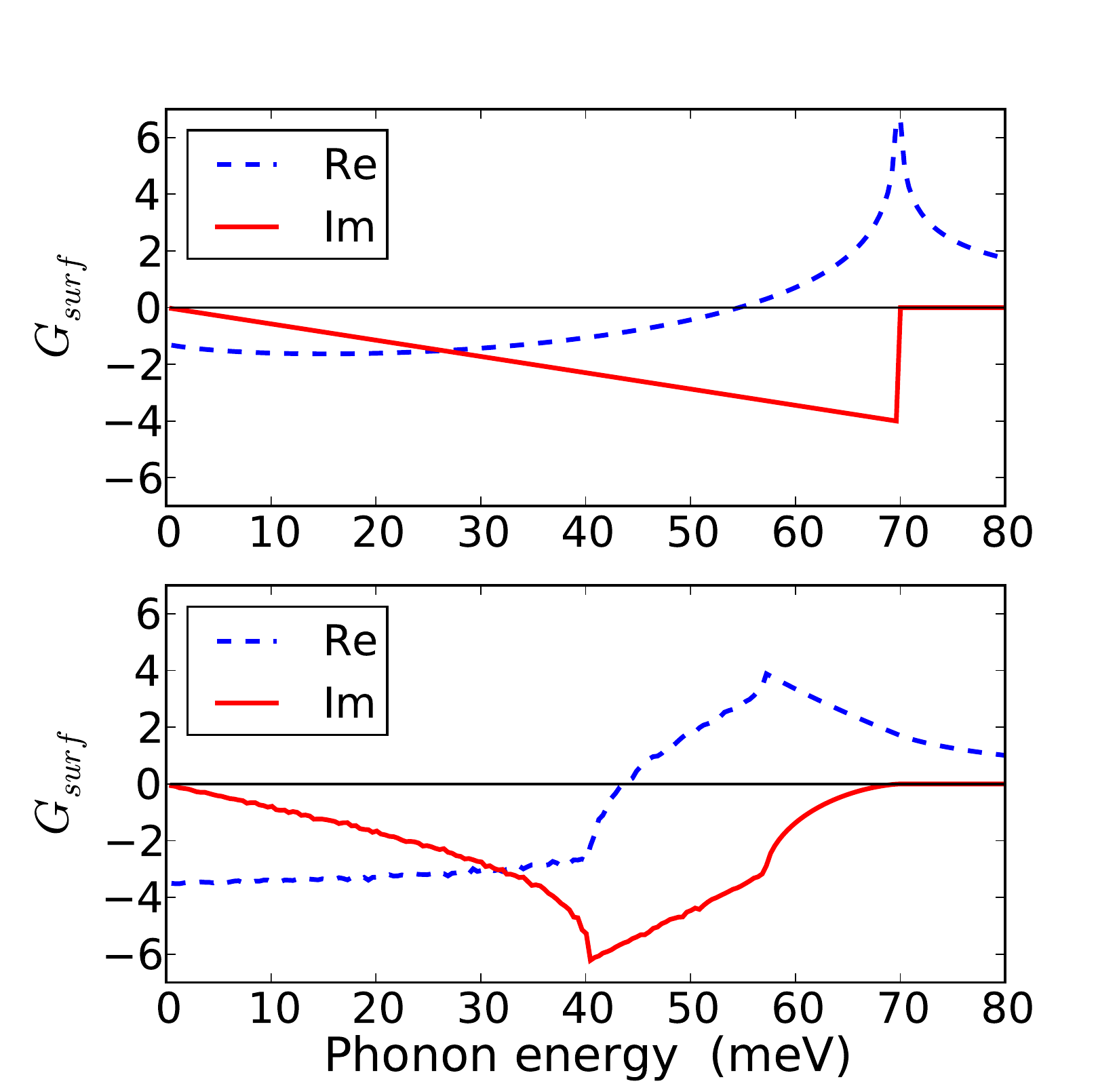}
\caption{Green's function of a bare surface atom without coupling to the molecule. Panel (a) show the analytical model and panel (b) show the result for a simple cubic crystal. }
\label{surface_GF}
\end{figure}

Figure \ref{surface_GF} (a) illustrate the shape of the bare surface Green's function (GF) from the analytical model that we have adopted from Ref. \onlinecite{Mingo2006}. For comparison we show in panel (b) the surface GF obtained numerically for a simple, isotropic cubic lattice with only nearest neighbor force constants. In both cases the maximum phonon energy is $E_{ph}^{max}=\hbar\omega_D=70\,$meV. For the simple cubic lattice we use a mass-scaled force constant between neighboring atoms of $k=-\omega_D^2/12$. In both cases, we have $\frac{-1}{\pi}\int_0^{\omega_D}{\rm Im}G(\omega)d(\omega^2)=1$ corresponding to one degree of freedom. 

\section{Derivation of transmission zeros for meta-benzene} \label{graphical}
\begin{figure}[htb!]
\includegraphics[width=0.8\columnwidth]{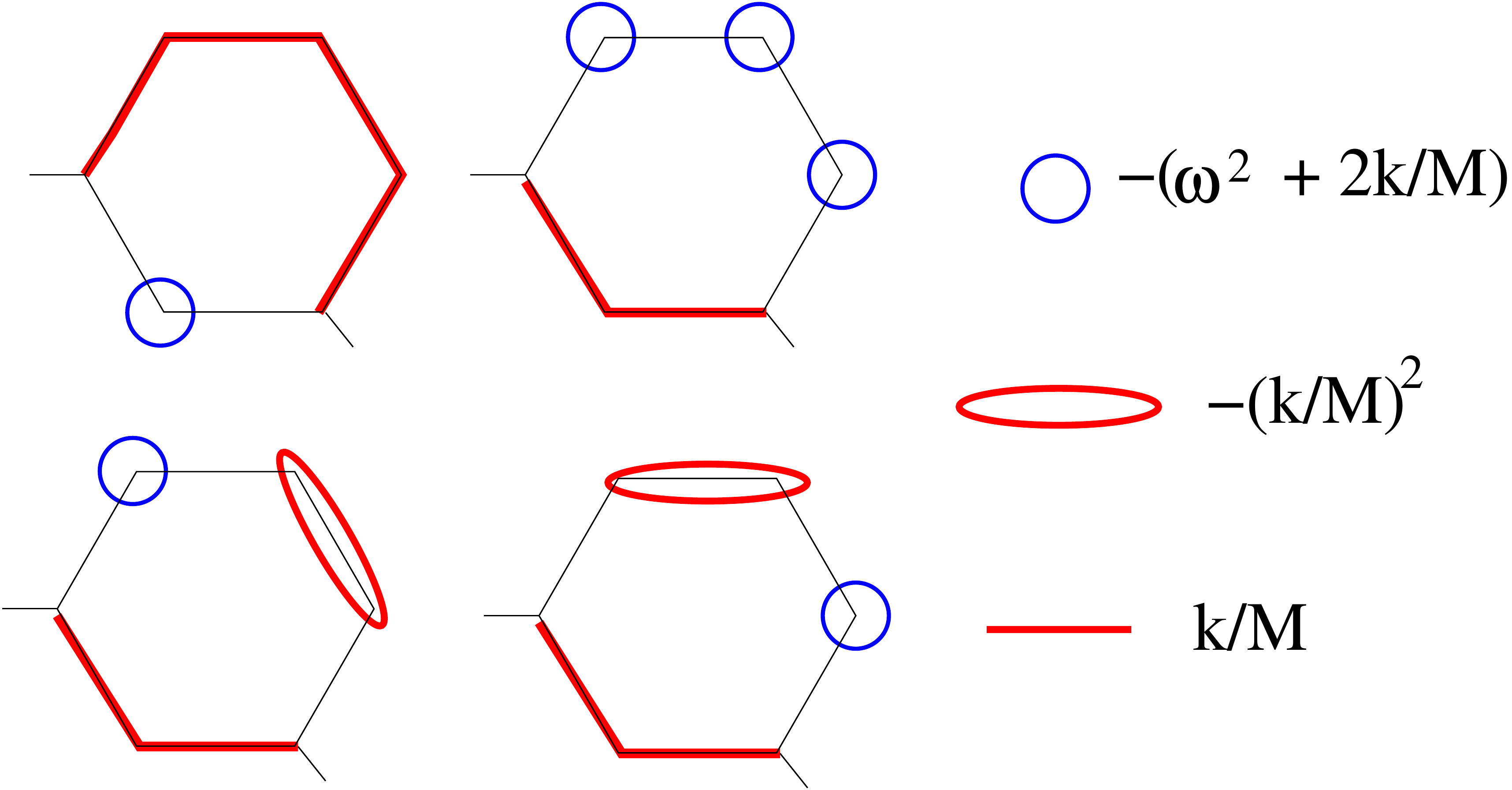}
\caption{Graphical representation of the (1,3) cofactor of $(\omega^2-\mathbf{D})$ for meta-benzene in a nearest neighbor model considering only out-of-plane motion. Each diagram represents a term in the determinant. To the right is shown the algebraic values corresponding to the different symbols, which should be multiplied for each diagram.}
\label{graphical_rules}
\end{figure}

In order to analyze the phonon transmission function in benzene, we apply a previously developed graphical scheme for predicting transmission zeros for electron transport. While details of this approach can be found in Ref. \onlinecite{Markussen2011}, we here briefly summarize the method. In a nearest-neighbor model, where the molecule is coupled to leads at site 1 and $N$, the phonon transmission function can be written as $\mathcal{T}(\omega)=\gamma(\omega)^2|G_{1N}(\omega)|^2$, where $\gamma$ includes the lead density of states and the coupling to the molecule, and $G_{1N}$ is the $(1,N)'$th element in the Green's function matrix. Applying Cramers rule, $G_{1N}$ is proportional to the $(1,N)$ cofactor of the $(\omega^2-\mathbf{D})$ matrix, defined as the determinant of the matrix obtained by removing the first row and the $N$'th column of $(\omega^2-\mathbf{D})$. This determinant can be represented graphically using the following rules: (i) Site 1 and $N$ must be connected by a continuous path. (ii) The remaining sites must either be paired with nearest neighbors or have an on-site loop. (iii) Sum up all possible ways of fulfilling (i) and (ii). Figure \ref{graphical_rules} illustrate the four possible diagrams for meta-benzene.

The algebraic equation, determining the frequencies at which the transmission is zeros, is obtained from Fig. \ref{graphical_rules}:
\begin{eqnarray}
-\left(\omega^2+\frac{2k}{M}\right)\left(\frac{k}{M}\right)^4 -\left(\omega^2+\frac{2k}{M}\right)^3\left(\frac{k}{M}\right)^2&&\nonumber\\
+2\left(\omega^2+\frac{2k}{M}\right)\left(\frac{k}{M}\right)^4&=&0, \label{eq:zeros}
\end{eqnarray}
where $k$ is the nearest neighbor force constant and $M$ is the carbon atomic mass. From Eq. \eqref{eq:zeros} we obtain three possible transmission zeros at phonon energies $E_{ph}=\hbar\sqrt{-2k/M}=61.3\,$meV and $E_{ph}=\hbar\sqrt{-2k/M \pm|k|/M}$ giving 43.4 eV and 75.1 eV, in agreement with the numerical results in Fig. \ref{benzene:fig} (bottom). For a comprehensive analytical treatment of the transmission nodes in benzene in the electronic case we refer to e.g. Ref. \onlinecite{Hansen2009}.

Recall from Sect. \ref{method} that the on-site elements of the molecule dynamical matrix at the sites connecting to the leads need to be changed in order to fulfill momentum conservation, e.g. $D_{11}\rightarrow D_{11}-\gamma/M$, where $\gamma$ is the molecule-lead coupling, $M$ is the carbon atomic mass.
Notice, however, that the on-site elements at the sites connecting molecule and leads do not enter in the diagrams (there can be no on-site loops at the connecting sites). This implies that the transmission node energies are independent on the coupling to the leads ($\gamma$). This is contrary to the transmission peaks which occur at the eigen energies in the presence of lead coupling (including lead self-energies). Within first order perturbation theory the eigen-energies (and hence the transmission peaks) are shifted by $\Delta\omega_i=\sqrt{-\gamma(u_1^2+u_N^2)/M}$,  where $u_{1,N}$ is the amplitude of the phonon mode at the connecting sites 1 and N. In the case of mode 1 for benzene in Fig. \ref{benzene:fig} we have $u_1=u_N=1/\sqrt{6}$, $\gamma=-4\,$eV/\AA$^2$ leading to $\hbar\Delta \omega_1=21\,$meV, in good agreement with the position of the first transmission peak in Fig. \ref{benzene:fig}, which in addition to the upward shift $\Delta\omega$ also includes a (smaller) downward shift due to the (negative) real part of the lead self-energy.

\section{Out-of-plane modes for OPE3} \label{para-modes}
\begin{figure}[htb!]
\includegraphics[width=\columnwidth]{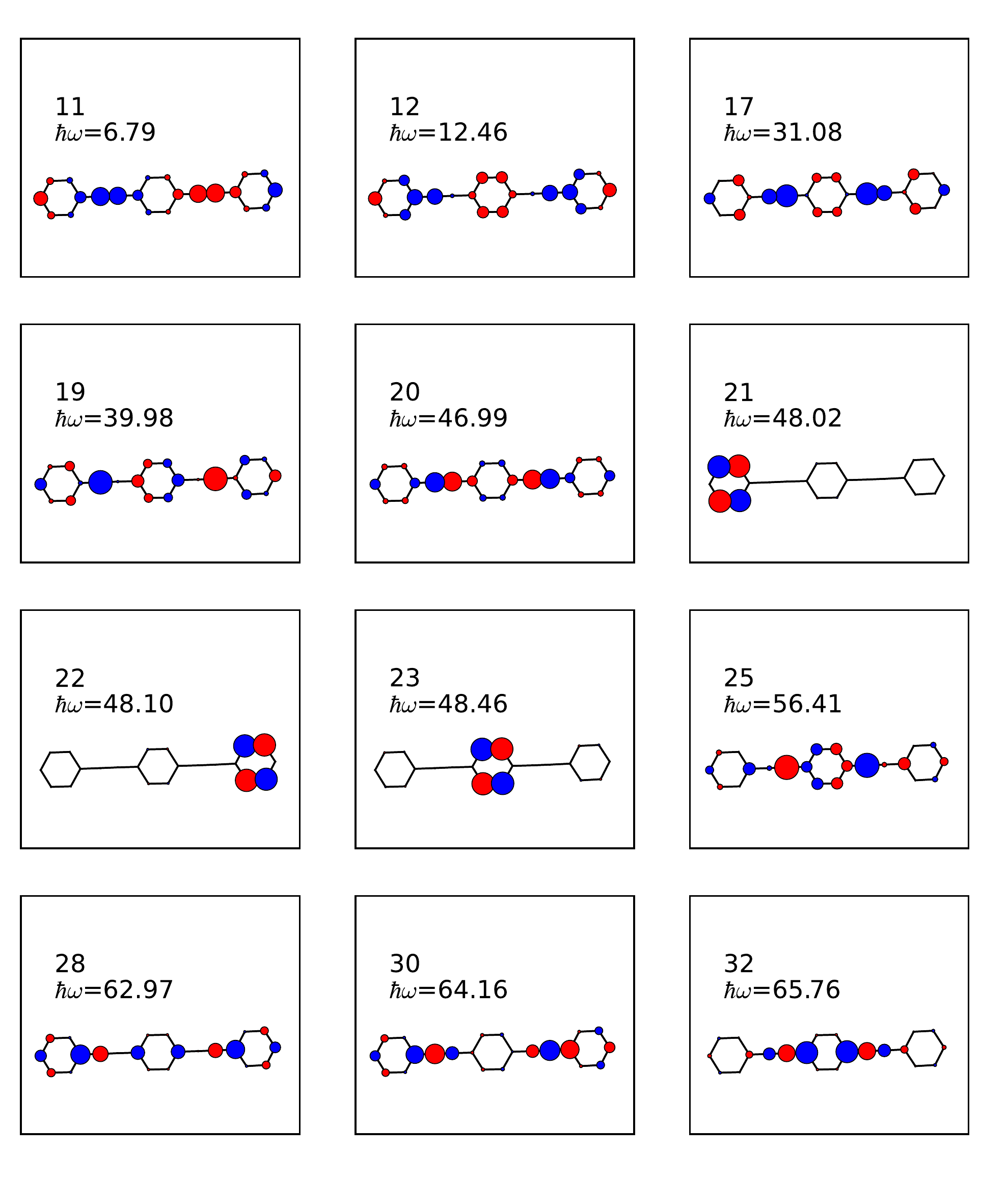}
\caption{Out-of-plane modes for para-OPE3 within the transport energy window. For each mode we indicate the mode index and the eigen-energy (in units of meV).}
\label{para-ope3-z-modes}
\end{figure}

Figure \ref{para-ope3-z-modes} shows the out-of-plane modes for the free para-OPE3. Additional modes occur at lower energies, but these do not contribute significantly to the transport. When the molecules are coupled to leads, the mode energies are shifted upward, but the shape of the modes are less affected.

\end{document}